\documentclass[12pt]{article}
\usepackage{amsmath,amsthm,amssymb,bm,mathrsfs}
\usepackage{graphicx}
\usepackage{typearea}
\typearea{12}

\newcommand{\Vc}[1]{\ensuremath{\mbox{\boldmath $#1$}}}
\newcommand{\tabtopsp}[1]{\vbox{\vbox to#1{}\vbox to1em{}}}

\title{Density Field of Dilute Particle Flow in Two-Fluid Model 
with Incompressible Carrier Flow}

\author{Kazuhiro TSUBOI
\thanks{Professor Emeritus, Ibaraki University.
Email: kazuhiro.tsuboi.508@vc.ibaraki.ac.jp}}

\date{August 2026}

\begin{document}
\maketitle

\begin{abstract}
We clarify some mathematical features of the density field of dilute particle flow 
on body surface in a two-fluid model; 
this model is commonly used to represent water droplet flows in icing simulations. 
In the case of incompressible carrier flows, 
a qualitative classification based on the Stokes number reveals 
four types of density fields of the dispersed phase. 
Inviscid solutions in the vicinity of a front stagnation point are obtained 
in exact and perturbed forms for small and large Stokes numbers, respectively. 
A comparison between the present solutions and numerical results shows 
good agreement and clarifies the behaviour of the density of the dispersed phase 
over the entire range of Stokes numbers. 
In addition, the viscous effect of the laminar boundary layer in the carrier phase 
on the particle-free thin layer appearing in the low Stokes number regime is 
investigated based on the incompressible Navier-Stokes equations. 
The order estimation of the velocity field of the dispersed phase reveals 
the cause and appearance criterion of this layer. 
These conclusions are supported by numerical results. 
In particular, the latter result on the criterion gives 
an improved form of Michael's criterion. 
\end{abstract}


\section{Introduction}

Several multiphase flow models have been proposed in order to represent flows 
including a large number of small particles. 
In these models, one of the simplest ways to represent 
the flows is a two-fluid description, 
in which the individual motion of small particles dispersed in a flow is averaged 
and the dynamical state of particle clouds is expressed 
in terms of some field quantities such as the density and the velocity\cite{Crowe}. 
This description is therefore valid only when the flow is dilute, 
that is, the effect of interaction between the dispersed particles is negligibly small. 
This condition can usually be measured by the bulk concentration of the dispersed phase. 

A basic equation of the dispersed phase in a two-fluid model, 
in particular, one with an incompressible carrier flow, was employed by Saffman 
to investigate the stability of dusty gas\cite{Saffman}, 
in which flows of a gas carrying small solid particles were considered. 
This equation includes a parameter called the Stokes number; 
the Stokes number is given as the non-dimensional relaxation time of particle motion 
to the characteristic time of the carrier flow. 
In this equation, the value of the Stokes number determines 
the behaviour of the flow field of the dispersed phase. 
Therefore, it is important to understand the dependency of the flow field 
on this parameter. 
After Saffman's work, some theoretical investigations 
on the flow field of a dispersed phase around an obstacle have been performed 
using an approximated form of the equation, 
in which the potential flow of the carrier phase was assumed. 
On the basis of this approach, 
some important features of the flow problem have been clarified
\cite{Michael_68,Michael_69,Healy_70a,Healy_70b,Healy_71,Sone}.\ 
These features are briefly summarized as follows.  

Here, we consider an obstacle in a uniform flow with dispersed particles. 
Since the effect of finite-sized suspended particles is not included 
in the abovementioned governing equation, 
there exits the threshold of the Stokes number in particle impingement 
to the body surface\cite{Marble}. 
Below the threshold, particle impingement on the body surface does not occur 
and particles flow downstream along the body surface. 
In this case, a thin layer of free particles appears along the surface, 
and the thickness of this layer is of the order of magnitude 
of the Stokes number\cite{Michael_68,Healy_70a,Healy_70b,Tsuboi_00}.
Simultaneously, a singularity appears in the particle density 
on the body surface\cite{Michael_68}.
 
On the other hand, beyond the threshold of the Stokes number, 
the uniform flow of the particle phase at upstream infinity cannot follow 
the flow of the carrier phase due to the large inertia, 
and particles impinge on the body surface\cite{Michael_69}.  
In this case, no particles are carried behind the body 
and the wake of the particle phase forms\cite{Sone}. 

However, the behaviour of the flow field of a dispersed phase has not yet been clarified. 
In fact, interesting results have been obtained 
only when the Stokes number is small or large sufficiently. 
For example, the dependency of the flow field on the wide range of the Stokes number 
has not yet been clarified although it is important knowledge 
in the dynamics of the dispersed particle flow. 
In particular, the singularity in the density of the dispersed phase 
strongly affects the application of the equation to some practical problems. 
In addition, the effect of the viscosity of the carrier flow 
on the particle-free thin layer needs further investigation. 
On the basis of a simple comparison between the thickness of the laminar boundary layer 
in carrier phase and that of the particle-free layer, 
Michael proposed a criterion for the appearance of this layer, 
$\tau \mathrm{Re}^{0.5} \gg 1$ where $\tau$ and Re indicate the Stokes number 
and the Reynolds number, respectively\cite{Michael_68}. 
However, a numerical result shows that this thin layer appears 
under the condition $\tau \mathrm{Re}^{0.5} \sim \mathrm{O}(1)$\cite{Tsuboi_04}. 

In ice accretion phenomenon, supercooled water droplets impinge and accrete 
as ice on structures; 
this phenomenon is observed in artificial structures such as aircraft\cite{Gent} 
or in natural circumstances\cite{Phillips}. 
Such an airflow with water droplets is a typical example of a flow 
with a small bulk concentration, 
since the mass of water in a unit volume of air, 
called the liquid water content (LWC)\cite{Launiainen}, 
is rather small, with the value of LWC being at most $1$ g/m³ in general\cite{Makkonen}. 
In such a case, it is obvious that collisions between dispersed water droplets 
can be neglected. 
Thus, the two-fluid model has been widely used to represent atmospheric flows 
with supercooled water droplets in the prediction of icing problems. 
In this context, the same equation as that used for investigating dusty gas 
can be applied to the basic equation of airflows containing water droplets.

In the 1980s, the two-fluid model was applied to simulations of ice accretion 
on a body surface, and numerical results of the local impingement efficiency 
along a surface were reported\cite{McComber,Scott,Bourgault};
this is defined as the mass flux of oncoming water droplets at a surface, 
and it plays an important role in estimating the amount of accreted ice\cite{Poots}. 
Since the 1990s, simulated shapes of accreted ice have been obtained numerically 
and comparisons of the computed and experimental results 
have become possible\cite{Hedde,Naterer,Tsuboi_12,Yoon}.
However, the issues with the abovementioned basic equation for droplets 
have been unsolved in these works\cite{Tsuboi_07}. 

In this study, we analytically and computationally investigate 
the basic equation of dilute particle phase in the two-fluid model 
and clarify some mathematical features of the density field of the dispersed phase 
in the vicinity of a body. 
A qualitative classification of the density 
at the surface of an incompressible carrier flow is described. 
Analytical solutions of the density in the vicinity of a stagnation point
are obtained successfully based on the inviscid stagnation flow model of a carrier phase, 
and the behaviour of the density is clarified 
over the entire range of values of the Stokes number. 
In addition, the viscous effect of the laminar flow of the carrier phase 
on a particle-free layer is also clarified 
and an improved criterion for this layer is proposed 
based on an order estimation and a numerical study 
using the incompressible Navier-Stokes equations.

\section{Summary of Basic Equation}

In this section, we summarize the basic equation of the dispersed phase in the two-fluid model 
and discuss some features of the basic equations 
in order to clarify the standpoint of the present investigation. 

\subsection{Basic Equation of Dispersed Phase}

On the basis of the two-fluid description of a multiphase flow, 
flows of dispersed particles are represented as field quantities of the density 
and the velocity for the dispersed phase as well as the carrier phase\cite{Crowe,Saffman}. 
Then, we consider flow problems around an obstacle placed in a uniform flow 
including a large number of dispersed particles 
moving with constant velocity $U_\infty$ 
and the density of the dispersed phase $\eta_\infty$ at upstream infinity.

When collisions between the particles are neglected and Stokes' drag formula 
is employed to model the interaction of the dispersed and carrier phases, 
the governing equations of the dispersed phase in the non-dimensional form 
are given as follows in the tensor notation: 
\begin{align} 
\label{eq:1} 
\frac{\partial \eta}{\partial t} + \frac{\partial\left(\eta v_j \right)}{\partial x_j} &= 0, \\
\label{eq:2}
\frac{\partial v_i}{\partial t} + v_j \frac{\partial v_i}{\partial x_j} 
+ \frac{1}{\tau} \left(v_i - u_i \right)  &= 0, 
\end{align}
where $\eta$ and $v_i$ mean the density and velocity fields for the dispersed flow, 
and $u_i$ does the velocity field for the carrier flow. 
These quantities are non-dimensionalized by $\eta_\infty$ and $U_\infty$, respectively. 
For simplicity, the effect of mass change is neglected here. 

The parameter $\tau$ included in Eq.~\eqref{eq:2} is a coefficient of fluid resistance 
between a dispersed particle and the carrier flow, 
and it physically represents the relaxation time of particle motion. 
The non-dimensional parameter $\tau$ is called the (particle) Stokes number 
and it is defined as follows for a spherical particle:
\begin{equation}
\tau^{-1} \equiv 3 \pi \mu \frac{d}{m} \frac{L}{U_\infty},
\end{equation}
where $d$ and $m$ denote the averaged diameter and mass of a dispersed particle, 
respectively, and $\mu$, the viscosity of the carrier fluid. 

For example, the diameters of dispersed water droplets in natural wind 
range from a few micrometers to tens of micrometers\cite{Makkonen,Finstad}. 
Then, for the flow condition of $U_\infty = 10$ m/s and $L = 1$ m, 
the value of $\tau$ is approximately of the order of $10^{-2}$, implying 
that the relaxation time of water droplets in motion is generally estimated 
to be of the order of milliseconds. 

\subsection{Qualitative Classification of Dispersed Phase Density at Stagnation}

We estimate qualitatively the density of the dispersed phase $\eta$ 
in the vicinity of the body surface. 
In the stagnation region, Eq.~\eqref{eq:1} is approximated well 
by the following equation in a steady flow: 
\begin{equation}
\label{eq:4}
\frac{1}{\eta}\frac{\partial \eta}{\partial n} \sim 
\frac{1}{v_n}\frac{\partial v_j}{\partial x_j} = \frac{1}{v_n}\mathrm{div}\Vc{v},
\end{equation}
where $v_n$ denotes the normal velocity component of the dispersed phase 
on the body surface.  

Using this expression, we can qualitatively classify the behaviour of the density 
of the dispersed phase, as shown in Table \ref{tbl:1}. 
Here, $\tau_{cr}$ denotes the threshold of the Stokes number; 
below this threshold, particles do not impinge the body surface. 
For simplicity, we assume that the carrier flow $u_i$ is incompressible; 
this table can then be explained as given below. 

\begin{table}[htbp]
\caption{Qualitative classification of dispersed phase density at stagnation}
\centering\smallskip
\begin{tabular}{c|ccc|c}
\hline\hline \tabtopsp{1.0mm}%
\hspace{25mm}      & 
\hspace{6mm}div\Vc{v}\hspace{6mm} &
\hspace{8mm}$v_n$\hspace{8mm} &
\hspace{8mm}$\eta$\hspace{10mm} &
\hspace{8mm}case\hspace{8mm} \\[1.0mm] 
\hline\tabtopsp{1.0mm}%
$\tau = 0$        & $= 0$     & $= 0$   & const. & I \\[1.0mm]  
\tabtopsp{1.0mm}
$\tau<\tau_{cr}$  & $\neq 0$  & $= 0$   & $0$ or $\infty$ & II \\[1.0mm]  
\tabtopsp{1.0mm}
$\tau>\tau_{cr}$  & $\neq 0$ & $\neq 0$ & $< \infty$      & III \\[1.0mm]  
\tabtopsp{1.0mm}
$\tau = \infty$   & $= 0   $ & $\neq 0$ & const. & IV \\[1.5mm]  
\hline\hline
\end{tabular}
\label{tbl:1}
\end{table}
For $\tau = 0$ (case I), the flow field of the dispersed phase is completely 
the same as that of the carrier flow, as given by Eq.~\eqref{eq:2}, 
and thus, it becomes incompressible everywhere. 
Then, the density of the dispersed phase has a finite value $\eta_\infty$. 
When $\tau < \tau_{cr}$ (case II), dispersed particles cannot reach the surface, 
implying that the normal velocity of the dispersed phase $v_n$ vanishes. 
However, incompressibility is not satisfied in this parameter region 
due to the compressibility of the dispersed phase, 
which is expected from Eq.~\eqref{eq:1}; 
this case is described in detail in Sec. 3.2. 
As the result, the density could become zero or infinite from Eq.~\eqref{eq:4}. 

Conversely, particle impingement occurs and the normal velocity $v_n$ remains 
finite for $\tau > \tau_{cr}$ (case III). 
The flow of the dispersed phase is compressibile in this region. 
Therefore, it is possible that the density at the surface has a finite value. 
Finally, for $\tau = \infty$ (case IV), no exchange of momentum occurs 
between the dispersed and carrier phases, and Eq.~\eqref{eq:1} 
therefore reduces to a simple convective equation. 
Using this equation, a solution of the uniform distribution of the dispersed phase 
exists everywhere except behind the obstacle\cite{Sone}, 
and the uniform flow of the dispersed phase reaches the surface 
only at the upstream side. 
The normal velocity $v_n$ has a finite value 
while the incompressibility of the dispersed phase is obvious. 
Therefore, the density of the dispersed phase has a constant value of $\eta_\infty$ or $0$. 

\section{Inviscid Solutions of Dispersed Phase Density in Stagnation Region}

In this section, we obtain analytical solutions of the dispersed phase 
by solving the steady equations of Eqs.~\eqref{eq:1} and \eqref{eq:2} under the assumption 
that the carrier flow $u_i$ is an inviscid stagnation flow. 
Although inviscid solutions are not completely accurate in real situations, 
they help in understanding the local features of the dispersed phase 
in the vicinity of a stagnation region. 

\subsection{Stagnation Flow Model}

We consider an irrotional stagnation flow as a model of the incompressible flow 
of the carrier phase; it is given in a non-dimensional form as follows: 
\begin{equation} \label{eq:5}
\begin{cases}
u_1 = 2 A x_1, \\
u_2 = - \alpha A x_2,
\end{cases}
\end{equation}
where $\alpha$ takes a value of $2$ or $1$ for a plane or axi-symmetric flow, respectively. 
In addition, $A$ is a constant, and the negative value represents flows 
near a front stagnation while a positive value indicates 
that the flows simulate separation regions. 

The numerical values of $A$ and $\alpha$ are summarized in Table 2 
for obstacles having typical shapes. 
The values of the threshold $\tau_{cr}$ for each obstacle are also listed;
these are obtained by solving the one-dimensional motion equation of a particle \cite{Marble}. 

\begin{table}[htbp]
\caption{Numerical examples of $A$, $\alpha$, and $\tau_{cr}$}
\centering\medskip
\begin{tabular}{c|ccc} 
\hline\hline \tabtopsp{1.0mm}%
\hspace{5mm}Geometry\hspace{5mm} &
\hspace{8mm}$|A|$\hspace{8mm}    &
\hspace{8mm}$\alpha$\hspace{8mm} & 
\hspace{8mm}$\tau_{cr}$\hspace{8mm} \\[1.0mm] 
\hline\tabtopsp{1.0mm}%
cylinder    & $2$     & $2$      & $1/16$      \\[1.0mm]  
\tabtopsp{1.0mm}
plate       & $1$     & $2$      & $1/8$       \\[1.0mm]  
\tabtopsp{1.0mm}
sphere      & $3$     & $1$      & $1/24$      \\[1.0mm]  
\tabtopsp{1.0mm}
disc        & $4/\pi$ & $1$      & $\pi/32$    \\[1.0mm]  
\hline\hline
\end{tabular}
\label{tbl:2}
\end{table}

\begin{figure}[htbp]
\centering
\includegraphics[width=.45\linewidth]{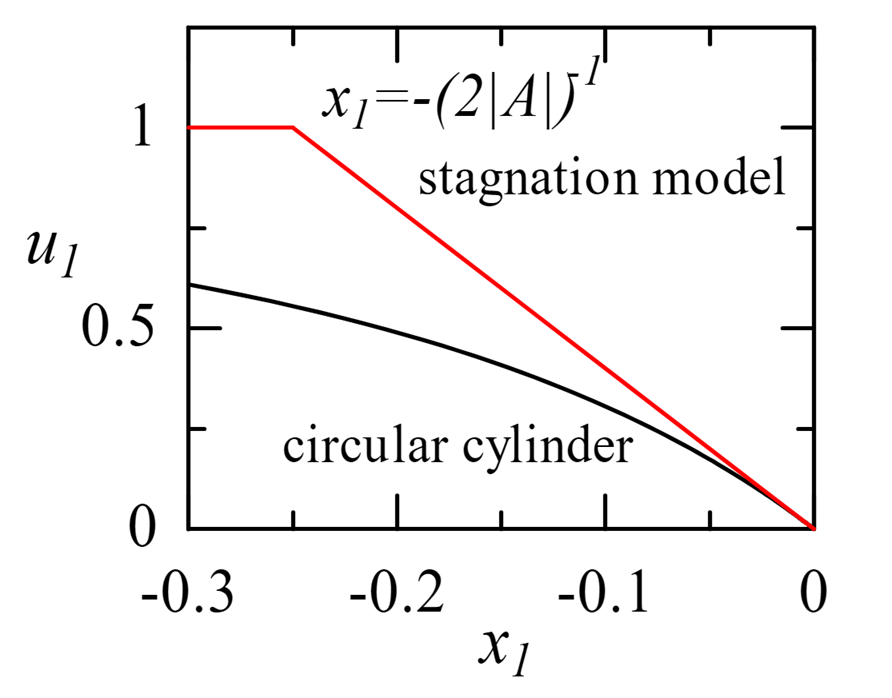}
\caption{Velocity profile of stagnation flow model \eqref{eq:5} along $x_1$-axis.}
\label{fig:1}
\end{figure}

The stagnation flow model \eqref{eq:5} is shown in Figure~\ref{fig:1}, 
in which the velocity $u_1$ on the $x_1$-axis (flow direction) is compared 
with that of the potential flow around a circular cylinder. 
In this model, the velocity becomes equivalent to the uniform velocity 
at $x_1 = - (2|A|)^{-1}$; this is equal to $-1/4$ in the case of a circular cylinder. 

\subsection{Exact Solution for Small Stokes Number}

When Stokes number $\tau$ is sufficiently small (case II in Table 1), 
the dispersed particles do not impinge on the body surface. 
In this parameter region, it is possible to obtain an exact solution 
since the carrier flow field $u_i$ is linear in the present stagnation model. 

In fact, we have a solution of the form 
\begin{equation} \label{eq:6}
\begin{cases}
v_1 = f(\tau) u_1, \\
v_2 = g(\tau) u_2,
\end{cases}
\end{equation}
where $f(\tau)$ and $g(\tau)$ are functions of only the Stokes number.
 
Substituting Eq.~\eqref{eq:6} into the $x_1$-component of Eq.~\eqref{eq:2}, 
we obtain the following algebraic equation for the function $f(\tau)$:
\[ 2 A \tau f^2 + f - 1 = 0. \]

Then, the solution of this equation is given as 
\begin{equation} \label{eq:7}
f(\tau) = \frac{-1 + \sqrt{1 + 8 A \tau}}{4 A \tau}, 
\end{equation}
where we choose $f(\tau)$ such that the solution approaches unity when $\tau \to 0$. 

Similarly, the $x_2$-component of Eq.~\eqref{eq:2} leads to the following solution 
of $g(\tau)$:
\begin{equation} \label{eq:8}
g(\tau) = \frac{1 - \sqrt{1 - 4 \alpha A \tau}}{2 \alpha A \tau}.
\end{equation}
 
The solutions $f(\tau)$ and $g(\tau)$ are depicted in Figure~2 
as functions of the Stokes number $\tau$. 
We can find that the functions $f(\tau)$ for $A < 0$ 
and $A > 0$ coincide with $g(\tau)$ for $A > 0$ and $A < 0$ 
in the case of a plane flow, respectively. 
\begin{figure}[htbp]
\begin{minipage}{.45\linewidth}
\includegraphics[width=\linewidth]{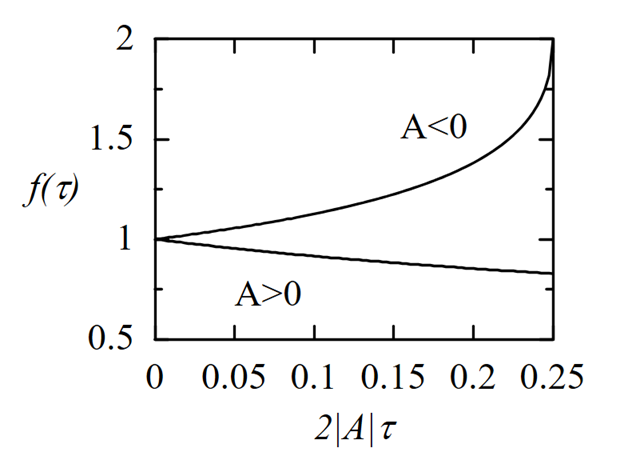}
\begin{center}
a) Profile of factor $f(\tau)$
\end{center}\label{fig:2a}
\end{minipage}
\begin{minipage}{.45\linewidth}
\includegraphics[width=\linewidth]{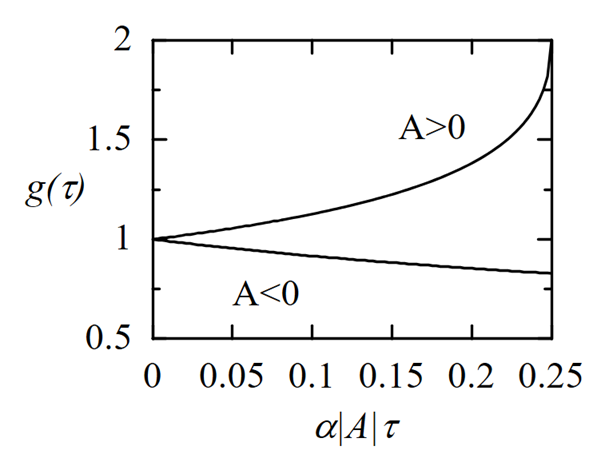}
\begin{center}
b) Profile of factor $g(\tau)$
\end{center}\label{fig:2b}
\end{minipage}
\caption{Factors of velocity field as functions of the Stokes number.}
\label{fig:2}
\end{figure}

Also, it is obvious that the similar solutions are valid in the following range 
of the Stokes number. 
\begin{align*}
0 &\le A \tau \le \frac{1}{4 \alpha},\quad A > 0, \\
0 &\le |A| \tau \le \frac{1}{8}, \quad A < 0.
\end{align*}

In particular, it should be noted that the upper limit of 
the range in $A < 0$ corresponds to the threshold of particle collision $\tau_{cr}$, 
as given in Table 2. 

Here, we consider the compressibility of the flow field of the dispersed phase. 
The divergence of the plane or axisymmetric flow of the dispersed phase is written as 
\[ 
\mathrm{div} \Vc{v} = \frac{\partial v_1}{\partial x_1}
+ \frac{1}{x_{2}^{2-\alpha}}\frac{\partial}{\partial x_2} \left( x_{2}^{2-\alpha} v_2 \right).
\]

Substituting solution \eqref{eq:6} into the above expression, we obtain
\begin{equation} \label{eq:9}
\mathrm{div} \Vc{v} = 2 A f - \alpha(3-\alpha) A g = 2 A \left( f - g \right),
\end{equation}
where the last expression is independent of the value of $\alpha$. 

Since it is found in Figure~\ref{fig:2} that $f - g \ge 0$ for $A < 0$ 
and $f - g \le 0$ for $A > 0$, the value of $\mathrm{div}\Vc{v}$ becomes 
negative irrespective of the value of $A$.  
Therefore, this result indicates that compressibility appears 
in the flow of the dispersed phase 
and its effect becomes increases with the value of $\tau$.  

Now, the density of the dispersed phase can be obtained directly 
from the velocity field. 
The density distribution $\eta$ is determined from Eqs.~\eqref{eq:1} 
and \eqref{eq:6} using Eq.~\eqref{eq:9}, and it is given as  
\begin{equation} \label{eq:10}
\eta = |x_1|^{\alpha(3-\alpha)g/(2f)-1}F\left( |x_1|^{\alpha g} |x_2|^{2f} \right),
\end{equation}
where $F$ is an arbitrary function that should be determined 
by the boundary condition. 

Since the curves $|x_1|^{\alpha g} |x_2|^{2f} = \mathrm{const.}$\ give 
the trajectories of dispersed particles, 
the value of $F$ remains constant along each trajectory. 
In particular, $F = 1$ is satisfied along $x_2 = 0$ 
because $\eta \to1$ when $\tau \to 0$.

It should be noted that solution \eqref{eq:10} is exact 
in the abovementioned range of the Stokes number. 
Considering the density distribution along a trajectory, 
we can clarify the singularity of the density field. 
In fact, we obtain the following relation for each trajectory 
for $\alpha = 1$ or $2$.
\begin{equation} \label{eq:11}
\eta \propto |x_1|^{g/f - 1}.
\end{equation}

The result shown in Figure~\ref{fig:2} indicates that $g/f - 1 > 0$ for $A > 0$ 
and $g/f - 1 < 0$ for $A < 0$, suggesting that the density of the dispersed phase 
in rear stagnation (or separation point) vanishes for a finite Stokes number 
while a singularity appears in the front stagnation. 
In order to clarify the singularity of Eq.~\eqref{eq:11}, 
we assume that the Stokes number is sufficiently small. 
Then, from expressions \eqref{eq:7} and \eqref{eq:8}, 
the density distribution can be reduced to 
\begin{alignat*}{3}
\eta &\sim |x_1|^{(\alpha+2)A \tau}, &\quad& A &> 0, \\
\eta &\sim \left(\frac{1}{|x_1|}\right)^{(\alpha+2)|A| \tau}, && A &< 0,
\end{alignat*}
according to the sign of $A$. 
The latter solution shows that the density in the front stagnation region 
increases with $|x_1|^{-\tau}$. 
In particular, the exponent of this singularity becomes $9\tau$ 
in the case of a sphere, 
which is in exact agreement with the result obtained by Michael \cite{Michael_68}.  

It should be noted here that solution \eqref{eq:10} does not describe 
a thin layer along a body surface, 
in which the density of the dispersed phase vanishes. 
This result is supported by numerical investigations on a dusty flow 
along a wedge with an arbitrary angle \cite{Healy_71}. 
This fact suggests that the centrifugal force of the surface curvature 
generates the particle-free thin layer 
that appears in the flow field of low Stokes numbers, 
which we will investigate in Sec. 4.  

\subsection{Perturbed Solution for Large Stokes Number} 

For $\tau > \tau_{cr}$ (case III in Table 1), 
each dispersed particle can reach the body surface due to the inertial motion. 
This implies that a solution having the same form as Eq.~\eqref{eq:6} 
does not exist in this parameter range. 
Therefore, we consider the perturbation from the uniform flow 
that corresponds to an infinite Stokes number (case IV in Table 1), 
and assume solutions of the following form:  
\begin{equation}
v_i  = \sum_{k} \left( \frac{1}{\tau} \right)^k v_i^{(k)}, \qquad
\eta = \sum_{k} \left( \frac{1}{\tau} \right)^k \eta^{(k)}. 
\label{eq:12}
\end{equation}

Substituting expressions \eqref{eq:12} into Eqs.~\eqref{eq:1} and \eqref{eq:2}, 
linearized equations are derived for each order of the Stokes number.

\noindent\medskip
\underline{0-th order} 
\begin{subequations}
\begin{align}
\label{eq:13a}
v_j^{(0)}\frac{\partial v_i^{(0)}}{\partial x_j} &= 0, \\
\label{eq:13b}
\frac{\partial}{\partial x_j}\left( \eta^{(0)} v_j^{(0)} \right) &= 0.
\end{align}
\end{subequations}
\noindent
\underline{1-st order} 
\begin{subequations}
\begin{align}
\label{eq:14a}
v_j^{(0)}\frac{\partial v_i^{(1)}}{\partial x_j} 
+ v_j^{(1)}\frac{\partial v_i^{(0)}}{\partial x_j} &= u_i - v_i^{(0)}, \\
\label{eq:14b}
\frac{\partial}{\partial x_j}
\left( \eta^{(0)} v_j^{(1)} + \eta^{(1)} v_j^{(0)} \right) &= 0.
\end{align}
\end{subequations}
\noindent
\underline{2-nd order} 
\begin{subequations}
\begin{align}
\label{eq:15a}
   v_j^{(0)}\frac{\partial v_i^{(2)}}{\partial x_j} 
 + v_j^{(1)}\frac{\partial v_i^{(1)}}{\partial x_j} 
 + v_j^{(2)}\frac{\partial v_i^{(0)}}{\partial x_j} &= - v_i^{(1)}, \\
\label{eq:15b}
\frac{\partial}{\partial x_j}
\left( \eta^{(0)} v_j^{(2)} + \eta^{(1)} v_j^{(1)} + \eta^{(2)} v_j^{(0)} \right) &= 0.
\end{align}
\end{subequations}

From these equations, we can easily obtain
the following solutions for each order of the Stokes number $\tau$. 
\begin{align*}
v_1^{(0)} &= 1,& v_2^{(0)} &= 0,& \eta^{(0)} &= 1, \\
v_1^{(1)} &=  A\left( x_1 - \frac{1}{2A} \right)^2,& 
v_2^{(1)} &= -2A\left( x_1 - \frac{1}{2A} \right) x_2,& 
\eta^{(1)} &= 0,
\end{align*}
\begin{align*}
v_1^{(2)} &= -\frac{A^2}{2}\left( x_1 - \frac{1}{2A} \right)^3\left( x_1 + \frac{1}{6A} \right),\\
v_2^{(2)} &= -\frac{2}{3} A^2\left( x_1 - \frac{1}{2A} \right)^2\left( x_1 - \frac{2}{A} \right)x_2,\\
\eta^{(2)}&= \frac{2}{3} A^2\left( x_1 - \frac{1}{2A} \right)^4,
\end{align*}
where it should be noted that these solutions are valid 
only in the upstream side of an obstacle. 

\begin{figure}[hbtp]
\centering
\includegraphics[width=0.45\linewidth]{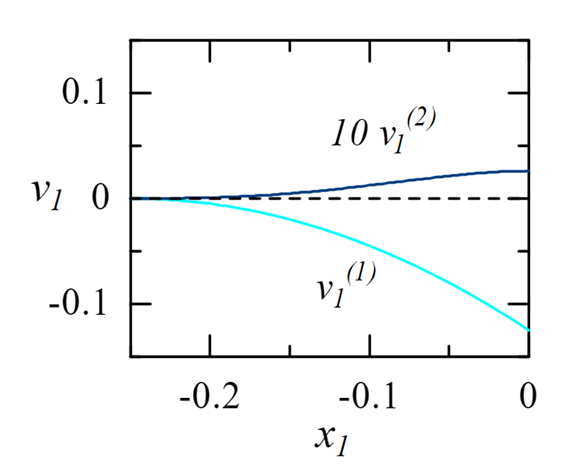}
\caption{Perturbed velocity profiles of $v_1^{(1)}$ and $v_1^{(2)}$ 
on the $x_1$-axis of a circular cylinder.}
\label{fig:3}
\end{figure}

The profiles of $v_1^{(1)}$ and $v_1^{(2)}$ on the $x_1$-axis are shown 
in Figure~\ref{fig:3} for a circular cylinder ($|A| = 2$). 
The second-order solution is sufficiently smaller than 
the first-order one indicating the effectiveness of the perturbed solution. 
In addition, we should note that the solution of the density 
in the first order $\eta^{(1)}$ vanishes identically. 
This fact holds for the series expansion of Eq.~\eqref{eq:12} in general \cite{Tsuboi_01}. 

The solution of the density of the dispersed phase is then obtained 
as follows: 
\begin{equation} \label{eq:16}
\eta = 1 + \frac{2}{3} A^2\left( x_1 - \frac{1}{2A} \right)^4\frac{1}{\tau^2}
+ \mathrm{O}\left(\frac{1}{\tau^3}\right).
\end{equation}
In the case of a circular cylinder, 
the density at the front stagnation is approximated as  
\begin{equation} \label{eq:17}
\eta = 1 + \frac{1}{96}\frac{1}{\tau^2} + \mathrm{O}\left(\frac{1}{\tau^3}\right).
\end{equation}

\subsection{Density of Dispersed Phase in Front Stagnation}

\begin{figure}[htbp]
\centering
\includegraphics[width=0.45\linewidth]{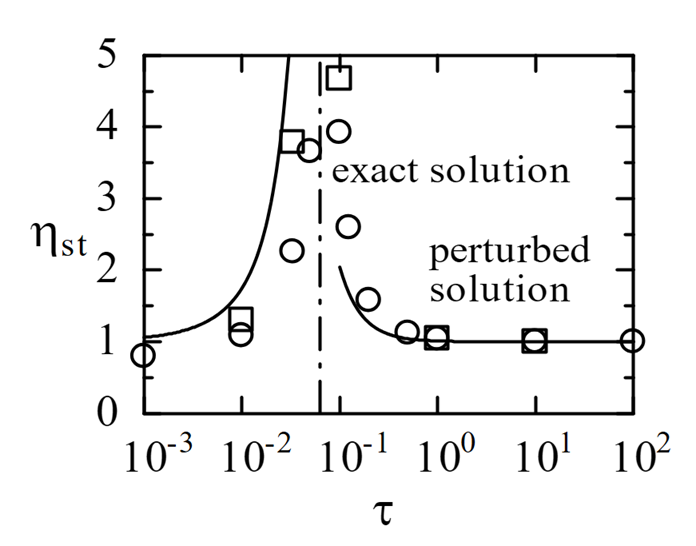}
\caption{Density of dispersed phase in front stagnation of a circular cylinder.}
\label{fig:4}
\end{figure}

In order to verify solutions \eqref{eq:11} and \eqref{eq:17}, 
a comparison with numerical solutions is performed. 
We employ the potential flow field around a circular cylinder 
as the flow of the carrier phase $u_i$ 
and the steady equations of Eqs.~\eqref{eq:1} and \eqref{eq:2} are solved numerically. 
The results of the density of the dispersed phase 
in the vicinity of the front stagnation are shown in Figure~\ref{fig:4}. 
In this figure, a dotted line is drawn 
at $\tau_{cr}$ ($=1/16$ in a circular cylinder); 
solution \eqref{eq:11} is valid to the left of this line 
while the perturbed solution \eqref{eq:17} is valid to its right. 
The computational results for two different grid resolutions are summarized: 
symbol $\circ$ corresponds to the results at $r/L \sim 7 \times 10^{-4}$ 
with $100 \times 70$ grid points 
and $\square$, those at $r/L \sim 8 \times 10^{-5}$ with $200 \times 200$ grid points, 
where $r$ and $L$ denote the distance from the front stagnation point 
and the diameter of a circular cylinder, respectively.  

The analytical and numerical results both show that the density 
has one peak near the threshold. 
Below the threshold, the density increases monotonously 
while the density decreases gradually above it. 
Therefore, we find that the features of both results qualitatively 
agree with each other. 
In particular, the perturbed solution is in good agreement with the numerical results 
although we employ a rather crude stagnation flow model, 
as shown in Figure~\ref{fig:1}.  

To the left from $\tau_{cr}$, a small difference is observed 
between the analytical and numerical results. 
This discrepancy is attributable to the boundary condition 
of numerical solutions. 
In the present computation, a simple Neumann boundary condition is employed 
at the cylinder surface. 
In this range of the Stokes number, the density has 
the singularity of $r^{-\tau}$ in the vicinity of the cylinder surface, 
leading to a drastic change in the density in this region. 
This implies that the resolution of the density profile is not sufficient. 
Since Eq.~\eqref{eq:2} is a hyperbolic equation, 
the boundary condition at the downstream region does not affect 
the upstream flow field seriously. 
However, we should focus on the underestimation of the density 
in the numerical solutions of this Stokes number region.

\section{Particle Free Layer and the Appearance Criterion}

Here, we consider an incompressible viscous carrier flow. 
In this case, it is known that a thin particle-free layer appears 
for a small Stokes number. 
On the basis of the order estimation in the basic equation 
and a numerical investigation, 
we obtain an improved criterion for the appearance of this layer.  

\subsection{Basic Equation of the Carrier Phase Flow}

We consider the Navier-Stokes equations for an incompressible flow 
as the governing equations of the carrier phase $u_i$ in Eq.~\eqref{eq:2}.  
\begin{align}
\label{eq:18}
\frac{\partial u_j}{\partial x_j} &= 0, \\
\label{eq:19}
\frac{\partial u_i}{\partial t} + u_j \frac{\partial u_i}{\partial x_j} 
&= -\frac{\partial p}{\partial x_i} 
 +  \frac{1}{\mathrm{Re}}\frac{\partial^2 u_i}{\partial x_j \partial x_j}
 +  \frac{\varepsilon}{\tau} \eta \left( v_i - u_i \right), 
\end{align}
where $p$ denotes the non-dimensional static pressure 
and $\mathrm{Re}$ does the Reynolds number of the carrier flow, respectively. 
In addition, the parameter $\varepsilon$ in the third term of the right-hand side 
in Eq.~\eqref{eq:19} denotes the density ratio of the dispersed and carrier phases, 
and it is given by  
\begin{equation}
\label{eq:20}
\varepsilon = \frac{\eta_\infty}{\rho_\infty},
\end{equation}
where $\rho_\infty\ (=\mathrm{const.})$ denotes the density of the carrier fluid. 

The third term on the right-hand side of Eq.~\eqref{eq:19} represents 
the reaction of fluid resistance from the dispersed phase to the carrier phase, 
and this effect is proportional to $\varepsilon$. 
Therefore, this term remains small as long as $\varepsilon/\tau$ is small. 
This assumption holds for dilute particle flows such as flows of dusty gas 
or supercooled fog, and therefore we can assume $\varepsilon < 1$ \cite{Saffman}.
In fact, the value of LWC in natural wind is at most $1\ \mathrm{g/m^3}$ 
as mentioned in Sec.~1, which means 
that the ratio of LWC to the density of air is of the order of $10^{-3}$ \cite{Tsuboi_12}. 
For the sake of simplicity, we assume $\varepsilon = 0.0$ here.  

Under this condition, the effect of the dispersed phase is negligible 
in the flow field of the carrier phase and the behaviour of the carrier flow is 
independent of the motion of dispersed particles. 
It is then possible to decompose the basic equations \eqref{eq:1}, \eqref{eq:2}, 
\eqref{eq:18} and \eqref{eq:19} into two subsystems of equations 
for the dispersed phase (Eqs.~\eqref{eq:1} and \eqref{eq:2}) 
and the carrier phase (Eqs.~\eqref{eq:18} and \eqref{eq:19}). 
Therefore, at every time step in the computation, 
the velocity field of the carrier phase $u_i$ is obtained using Eqs.~\eqref{eq:18} 
and \eqref{eq:19}, and then, the flow field of the dispersed phase follows 
the result of $u_i$ with Eqs.~\eqref{eq:1} and \eqref{eq:2}. 
Since we consider the steady flow fields in Sec.~4.3, 
time-averaged flow fields of the carrier phase are first computed 
using Eqs.~\eqref{eq:18} and \eqref{eq:19}, 
and those of the dispersed phase are obtained using Eqs.~\eqref{eq:1} and \eqref{eq:2} 
based on the time-averaged velocity field of $u_i$.  

\subsection{Order Estimation of Dispersed Phase for Low Stokes Number}

We consider the steady state of a dilute particle flow for a small Stokes number. 
Using Eq.~\eqref{eq:2}, the velocity field of the dispersed phase can be approximated 
as $v_i = u_i + \mathrm{O}(\tau)$ in this case. 
Substituting the expression into Eq.~\eqref{eq:2}, 
we obtain the following result as the first-order approximation: 
\begin{equation}
\label{eq:21}
v_i = u_i - \tau u_j \frac{\partial u_i}{\partial x_j} + \mathrm{O}(\tau^2)
   \approx u_i + \tau \left( \frac{\partial p}{\partial x_i} 
    -  \frac{1}{\mathrm{Re}}\frac{\partial^2 u_i}{\partial x_j \partial x_j} \right) 
    + \mathrm{O}(\tau^2), 
\end{equation}
where Eq.~\eqref{eq:19} is used to derive the last expression.  

Now, we introduce local coordinates along the tangential and normal directions 
of a streamline near a body surface, denoted by $s$ and $n$, respectively. 
We also assume the velocity profile in the laminar boundary layer as the carrier flow  
\[ 
u_s = \mathrm{O}(1),\qquad 
u_n = \mathrm{O}(\delta) \approx \mathrm{O}(1/\mathrm{Re}^{0.5}), 
\]
where $\delta$ denotes the thickness of the boundary layer, 
and the condition $\delta \ll 1$ is valid according to the boundary layer theory. 

The order of magnitude of the viscous terms in Eq.~\eqref{eq:21} is then 
estimated to be  
\[ 
\frac{1}{\mathrm{Re}}\frac{\partial^2 u_s}{\partial x_j \partial x_j} 
\approx \frac{1}{\mathrm{Re}}\frac{\partial^2 u_s}{\partial n^2} = \mathrm{O}(1),\qquad
\frac{1}{\mathrm{Re}}\frac{\partial^2 u_n}{\partial x_j \partial x_j} 
\approx \frac{1}{\mathrm{Re}}\frac{\partial^2 u_n}{\partial n^2} = \mathrm{O}(\delta), 
\]
and this estimation leads to the following expression of the tangential component. 
\begin{equation}
\label{eq:22}
v_s = u_s + \tau \left( \frac{\partial p}{\partial s} 
    - \frac{1}{\mathrm{Re}}\frac{\partial^2 u_s}{\partial n^2} \right)
    +\mathrm{O}(\tau^2). 
\end{equation}
Therefore, we have $v_s \sim u_s + \mathrm{O}(\tau)$ from Eq.~\eqref{eq:22}; 
this implies that the tangential velocity of the dispersed phase is 
of the same order of magnitude as the carrier phase. 

On the contrary, the estimation of the normal component is derived from Eq.~\eqref{eq:21} 
as follows: 
\begin{equation}
\label{eq:23}
v_n = u_n + \tau \left( \frac{\partial p}{\partial n} 
    - \frac{1}{\mathrm{Re}}\frac{\partial^2 u_n}{\partial n^2} \right) 
    + \mathrm{O}(\tau^2)
    \approx u_n + \tau\kappa v_{s}^2 
    + \mathrm{O}(\tau\delta) + \mathrm{O}(\tau^2),
\end{equation}
where $\kappa$ means the curvature of streamlines. 

The second term in the last expression of Eq.~\eqref{eq:23} is obtained 
from the following theorem \cite{Imai}: 
\begin{equation}
\label{eq:24}
\frac{\partial p}{\partial n} = \kappa u_s^2 
\approx \kappa v_s^2 + \mathrm{O}(\tau); 
\end{equation}
this implies that the centrifugal force due to the tangential velocity 
along a streamline with curvature balances the pressure gradient 
along the outward direction in steady inviscid flows. 
Moreover, the term of $\mathrm{O}(\tau\delta)$ in Eq.~\eqref{eq:23} originates 
from the viscous term in the normal direction. 
Since the order of magnitude of this term is estimated to be $\mathrm{O}(\delta^2)$ 
or $\mathrm{O}(\tau^2)$ depending on whether $\delta > \tau$ or $\delta < \tau$, 
this term then becomes a higher-order one for each case in the following discussion. 

Equation~\eqref{eq:23} shows that the order of the normal component $v_n$ 
depends on $\delta, \tau,$ and $\kappa$. When $\delta > \kappa\tau$, 
the normal velocity $v_n$ is comparable to $u_n$. 
In this case, the flow field of the dispersed phase is incompressible, 
as indicated by Eq.~\eqref{eq:18}, implying that no particle-free layer appears. 
However, the centrifugal effect becomes dominant in the normal velocity 
for $\delta < \kappa\tau$, because of which the flow of the dispersed phase 
deviates from the carrier flow. 
This effect causes the formation of a particle-free layer along a body surface, 
and the order of magnitude of $v_n$ is estimated to be $\mathrm{O}(\kappa\tau)$ 
for $\delta \ll \kappa\tau$. 

The above discussion leads to the following conclusions.  
\begin{enumerate} 
 \item Since the normal velocity of the dispersed phase due to the centrifugal force 
generates a particle-free layer, 
it appears along a surface or a streamline with curvature.  
 \item The particle-free layer appears under the condition $\delta < \kappa\tau$. 
In particular, the thickness of this layer is estimated to be $\mathrm{O}(\kappa\tau)$ 
when $\delta \ll \kappa\tau$.  
\end{enumerate} 

The former conclusion is supported by \cite{Healy_71} as well as the inviscid solution 
described in Sec.~3.2, and the latter provides an improved form of Michael's criterion 
mentioned in Sec.~1.  

\subsection{Numerical Results} 

Using the basic equations \eqref{eq:1}, \eqref{eq:2}, \eqref{eq:18} and \eqref{eq:19}, 
we numerically investigate the appearance of a particle-free layer. 
As mentioned in Sec.~4.1, the basic equations \eqref{eq:1}, \eqref{eq:2}, \eqref{eq:18} and \eqref{eq:19} 
are decomposed into two subsystems of Eqs.~\eqref{eq:1} and \eqref{eq:2}, and Eqs.~\eqref{eq:18} and \eqref{eq:19}, respectively. 
In the present computation, we consider flows around a circular cylinder, 
in which the Stokes number is $\tau = 0.02$ and the Reynolds number is varied. 
Equations \eqref{eq:1}, \eqref{eq:2}, \eqref{eq:18} and \eqref{eq:19} are discretized 
on a body-fitted coordinates system by using a finite difference method. 
In particular, the fine distribution of grid points near the body surface is used 
since a thin particle-free layer forms along the surface. 

Equations \eqref{eq:1} and \eqref{eq:2} for the dispersed phase are 
of the complete hyperbolic type. 
Therefore, in the discretization of the convective terms of these equations, 
a third-order upwind difference scheme is employed, 
in which the fourth-order numerical viscosity is added to the original term \cite{Kawamura}. 
In addition, the carrier flow is computed using the MAC algorithm 
with the Poisson equation for static pressure. 
In the discretization of Eq.~\eqref{eq:19}, 
the third-order upwind difference scheme is used in $\mathrm{Re} \ge 10^4$ 
for the same reason as that in Eqs.~\eqref{eq:1} and \eqref{eq:2}, 
while the computation in $\mathrm{Re} \le 10^3$ is possible 
using the central difference scheme having second-order accuracy \cite{Tsuboi_99}. 
In Eqs.~\eqref{eq:1} and \eqref{eq:2}, 
the absorbed condition of the dispersed phase is necessary 
at a body surface at the upstream side, 
by which dispersed particles can penetrate the surface, 
following which they are removed from the flow field. 
On the other hand, the no-penetration condition is imposed 
on the downstream surface.   

The computed profiles of the tangential velocity of the dispersed phase 
in the radial direction from the cylinder surface are shown in Figure~5 
for typical values of the Reynolds number. 
The solid and broken lines in each figure indicate the velocities of the dispersed 
and carrier phases, respectively. 
The profiles are given at several positions on the surface 
that are indicated by the angle measured from the front stagnation point. 
These results show that the velocity profile of the dispersed phase 
at each position is basically similar to that of the carrier phase. 
In fact, the two profiles are quite similar for $\mathrm{Re} = 10^2$. 
Therefore, we conclude that the velocity of the dispersed phase 
is of the same order of magnitude as that of the carrier phase, 
supporting the validity of the order estimation in Eq.~\eqref{eq:22}. 
\begin{figure}[hbtp]
\begin{minipage}{.45\linewidth}
\includegraphics[width=\linewidth]{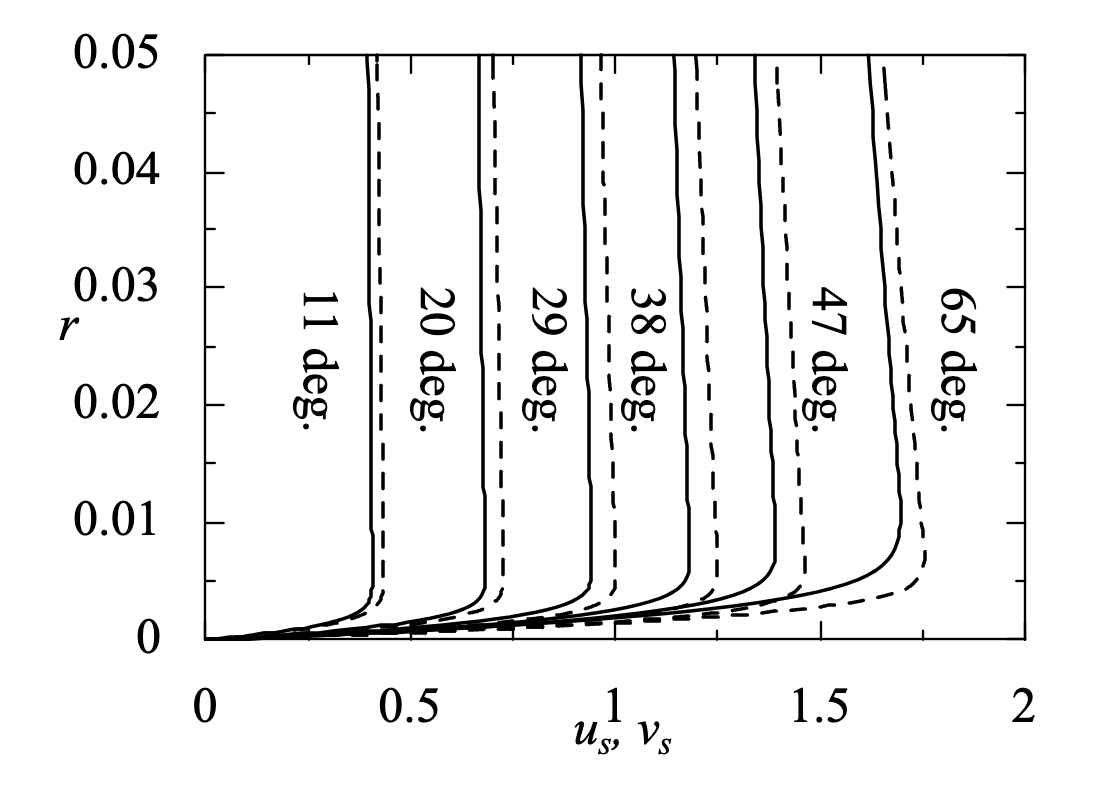}
\begin{center}
(a) Re $= 10^5$
\end{center}\label{fig:5a}
\includegraphics[width=\linewidth]{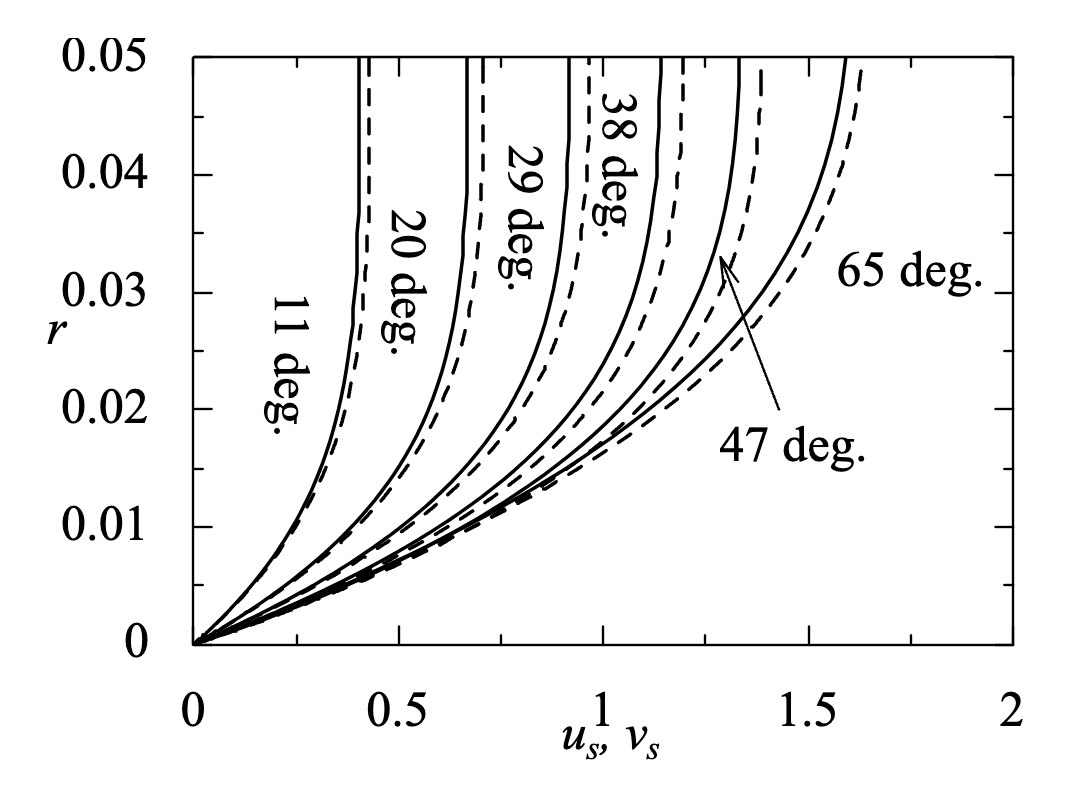}
\begin{center}
(c) Re $= 10^3$
\end{center}\label{fig:5c}
\end{minipage}
\begin{minipage}{.45\linewidth}
\includegraphics[width=\linewidth]{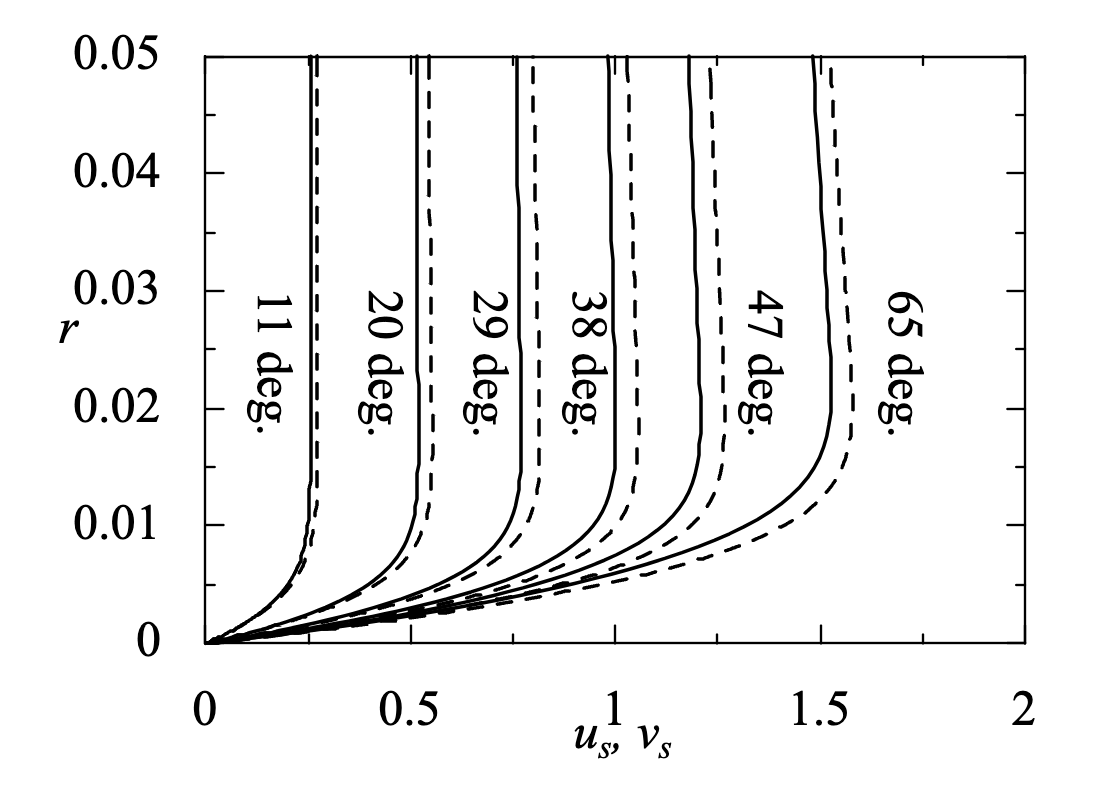}
\begin{center}
(b) Re $=10^4$
\end{center}\label{fig:5b}
\includegraphics[width=\linewidth]{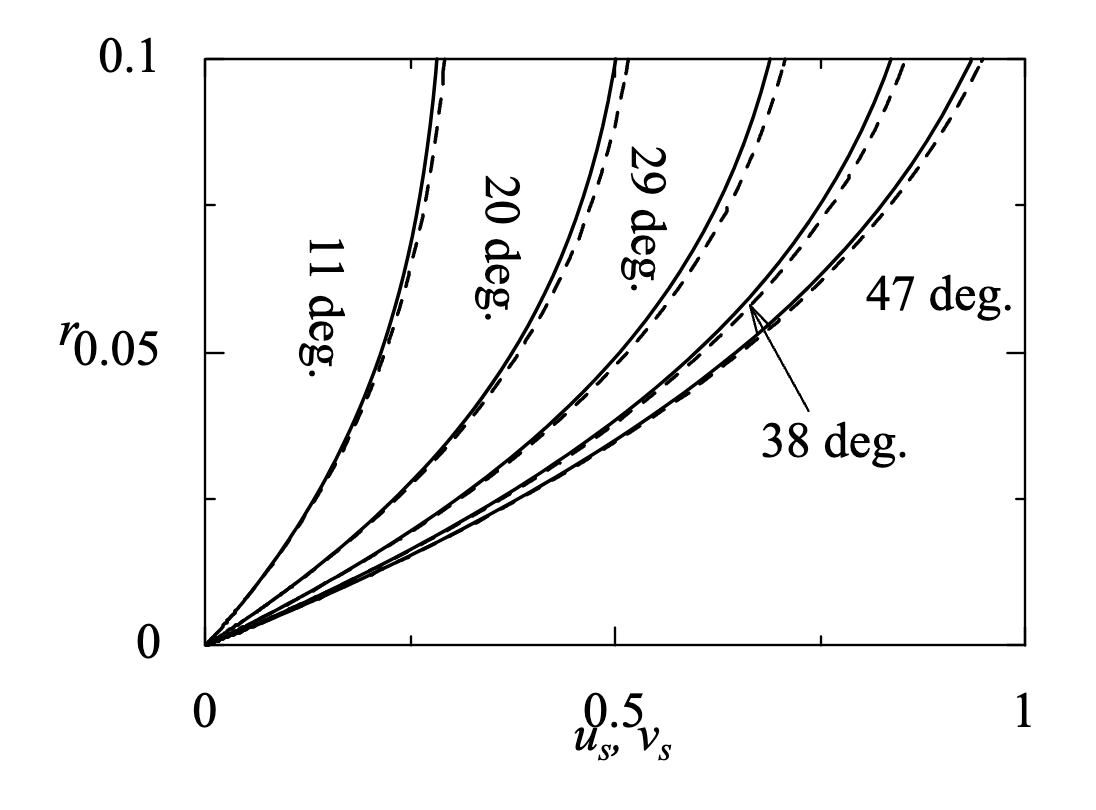}
\begin{center}
(d) Re $=10^2$
\end{center}\label{fig:5d}
\end{minipage}
\caption{Tangential velocity profiles of dispersed (solid line) 
and carrier (broken line) phases.}
\label{fig:5}
\end{figure}

We can summarize the tendency of the tangential velocity in the dispersed phase 
as follows. 
For the same Reynolds number, the tangential velocity of the dispersed phase 
becomes smaller than that of the carrier phase when moving downstream. 
This deceleration of the tangential velocity seems to be attributed to 
the outward motion of dispersed particles due to the normal velocity, 
and this expectation will be confirmed in the normal velocity profiles 
shown in Figure~6. 
In addition, the difference between the two profiles 
at the corresponding position increases with the Reynolds number. 
In particular, the maximum velocity increases and the location is closer 
to the cylinder surface with a higher Reynolds number. 
This implies that a larger centrifugal force is generated 
for a higher Reynolds number.  
\begin{figure}[htbp]
\begin{minipage}{.45\linewidth}
\includegraphics[width=\linewidth]{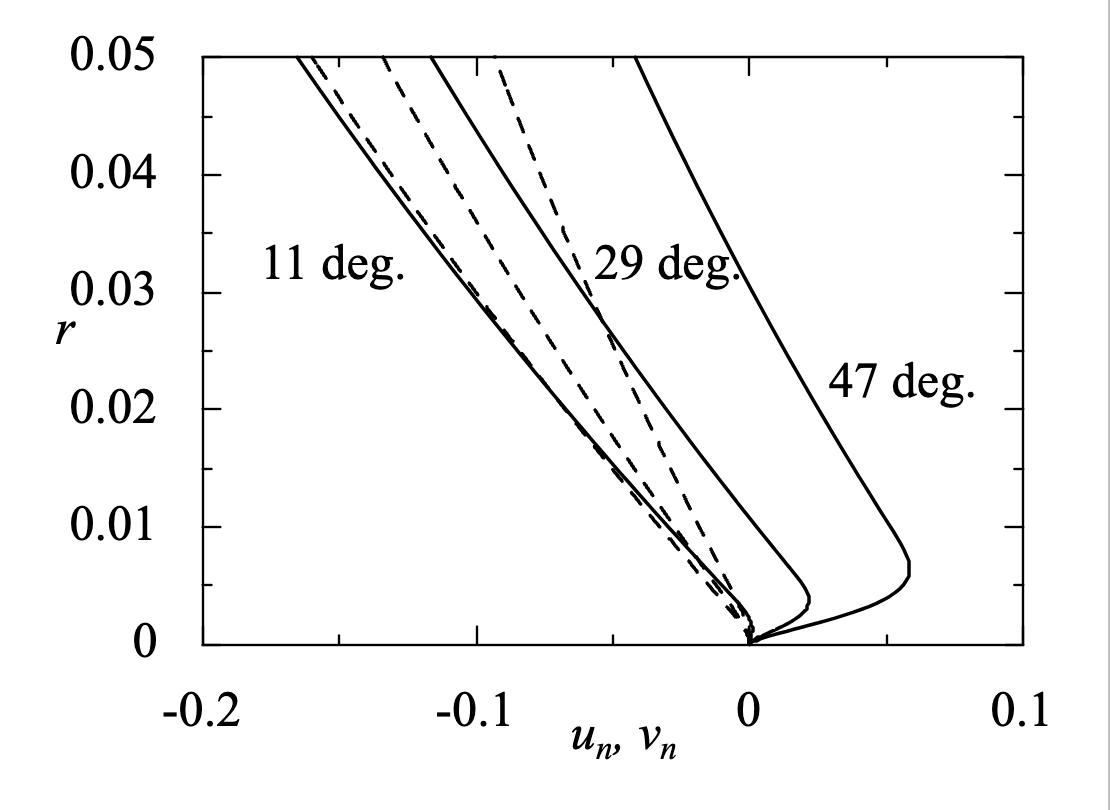}
\begin{center}
(a) Re $= 10^5$
\end{center}\label{fig:6a}
\includegraphics[width=\linewidth]{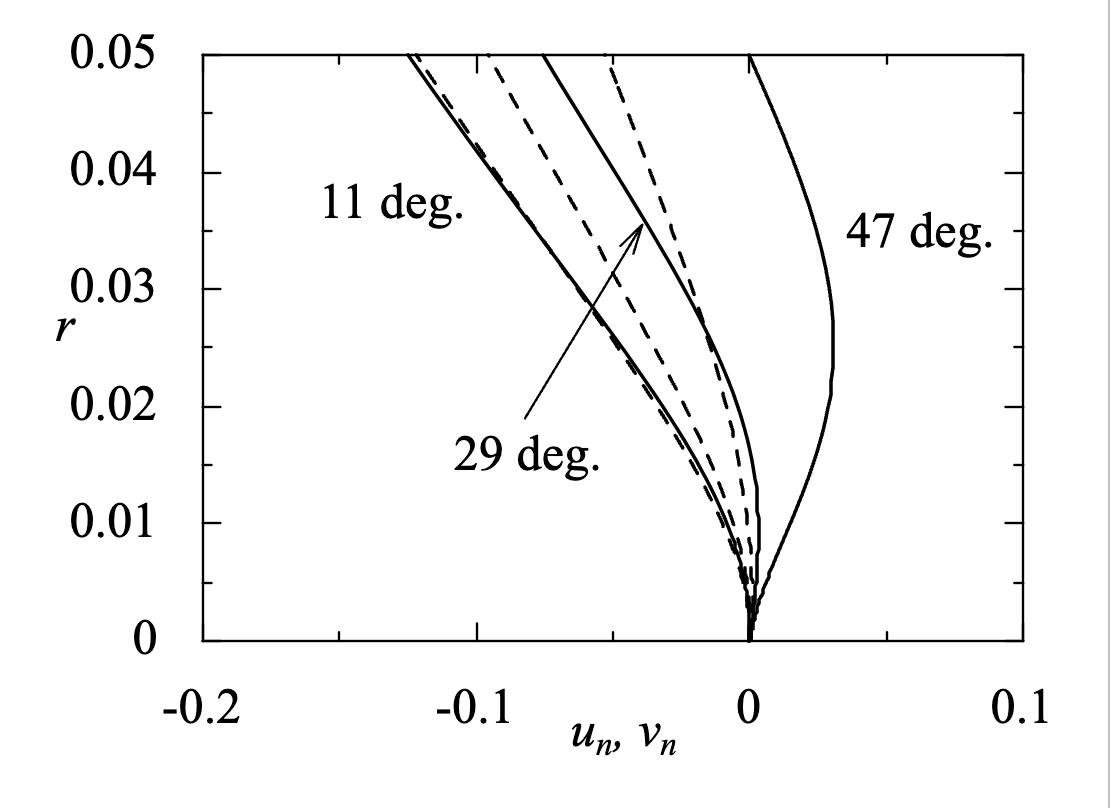}
\begin{center}
(c) Re $= 10^3$
\end{center}\label{fig:6c}
\end{minipage}
\begin{minipage}{.45\linewidth}
\includegraphics[width=\linewidth]{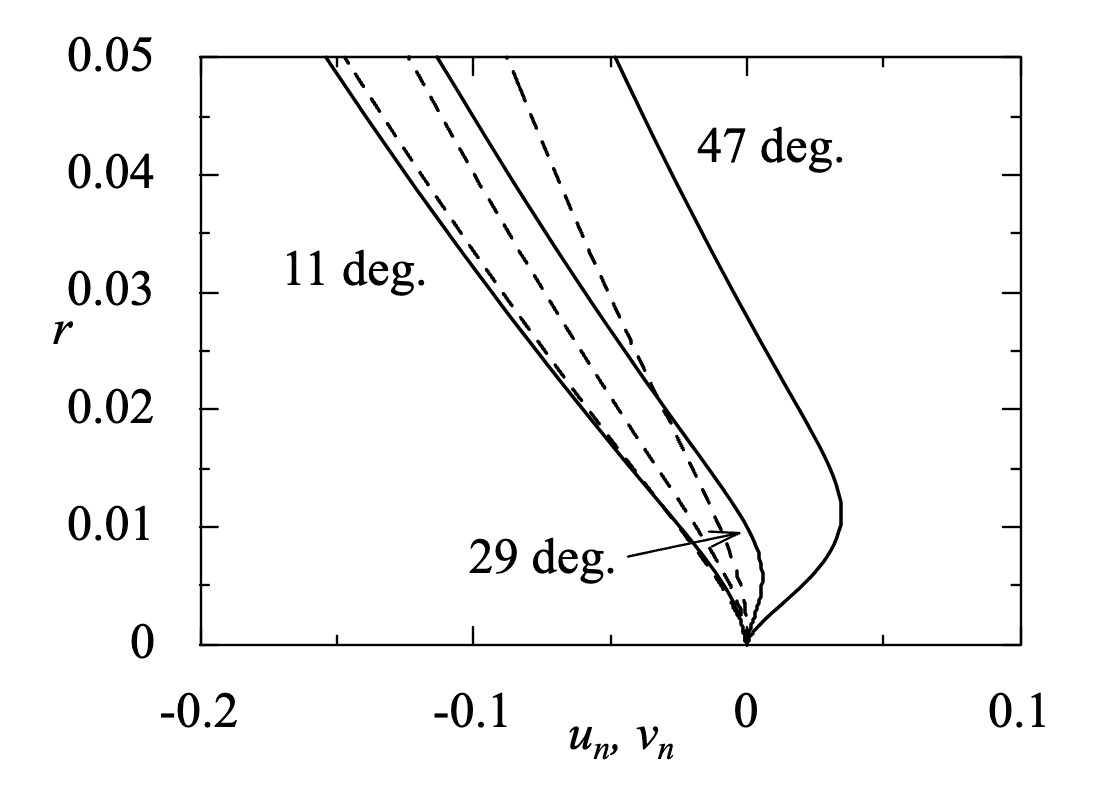}
\begin{center}
(b) Re $=10^4$
\end{center}\label{fig:6b}
\includegraphics[width=\linewidth]{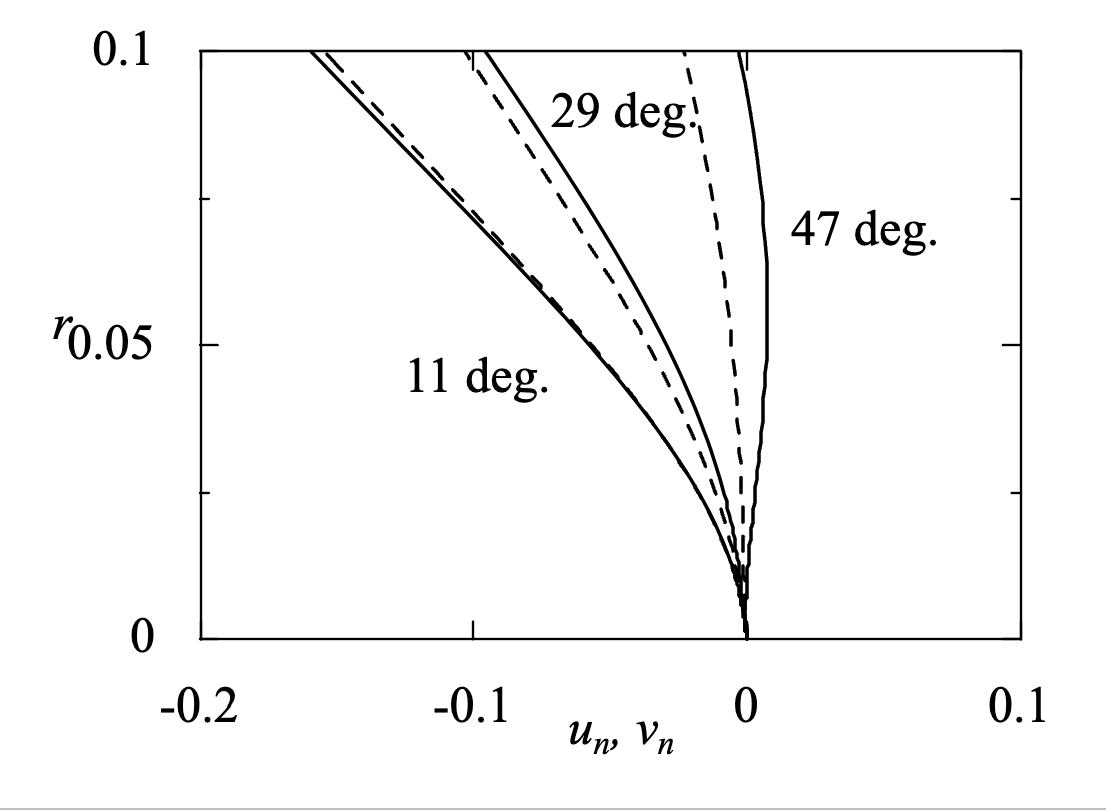}
\begin{center}
(d) Re $=10^2$
\end{center}\label{fig:6d}
\end{minipage}
\caption{Normal velocity profiles of dispersed (solid line) 
and carrier (broken line) phases.}
\label{fig:6}
\end{figure}

The normal velocity to the cylinder surface obtained in the present computation 
is shown in Figure~6 in a manner similar to that in Figure~5. 
Some remarkable results are observed in the results. 
We confirm that the order of magnitude of the normal velocity is much smaller than 
that of the tangential velocity, as assumed in the order estimation. 
The difference between the profiles of the dispersed and carrier phases 
is larger than that of the tangential velocity even for a low Reynolds number. 
In particular, the velocity of the dispersed phase in the downstream position 
is positive while that of the carrier phase is negative. 
This implies that the dispersed particles move away from the surface 
due to the centrifugal force generated by the tangential velocity. 
In fact, the location of the maximum normal velocity of the dispersed phase 
in each profile corresponds to the maximum region of the tangential velocity shown in Figure~5.  

The density profiles of the dispersed phase along the radial direction from the cylinder surface 
are shown in Figure~7. 
From the results, it is observed that the density $\eta$ vanishes 
in the finite range of $r$ for $\mathrm{Re} \ge 10^4$, 
indicating the formation of a particle-free layer. 
Furthermore, the particle-free layer develops downstream 
and its thickness is of the order of magnitude of the Stokes number ($\tau = 0.02$). 

On the other hand, when $\mathrm{Re} \le 10^3$, 
a low-density region of the dispersed phase occurs 
instead of the particle-free layer, 
that is, no particle-free layer appears in these cases. 
The reason for this result is evident. 
The appearance of the particle-free layer depends on 
the strength of the normal velocity of the dispersed phase. 
Particle trajectories separate from the surface 
due to the high normal velocity; 
in contrast, at a low velocity, 
a low-density region is formed instead of a particle-free region. 
When the Reynolds number is small (Figure~7 (c) and (d)), 
the density in the vicinity of the surface is found to increase drastically, 
which is explained as the singularity of the density field. 
\begin{figure}[htbp]
\begin{minipage}{.45\linewidth}
\includegraphics[width=\linewidth]{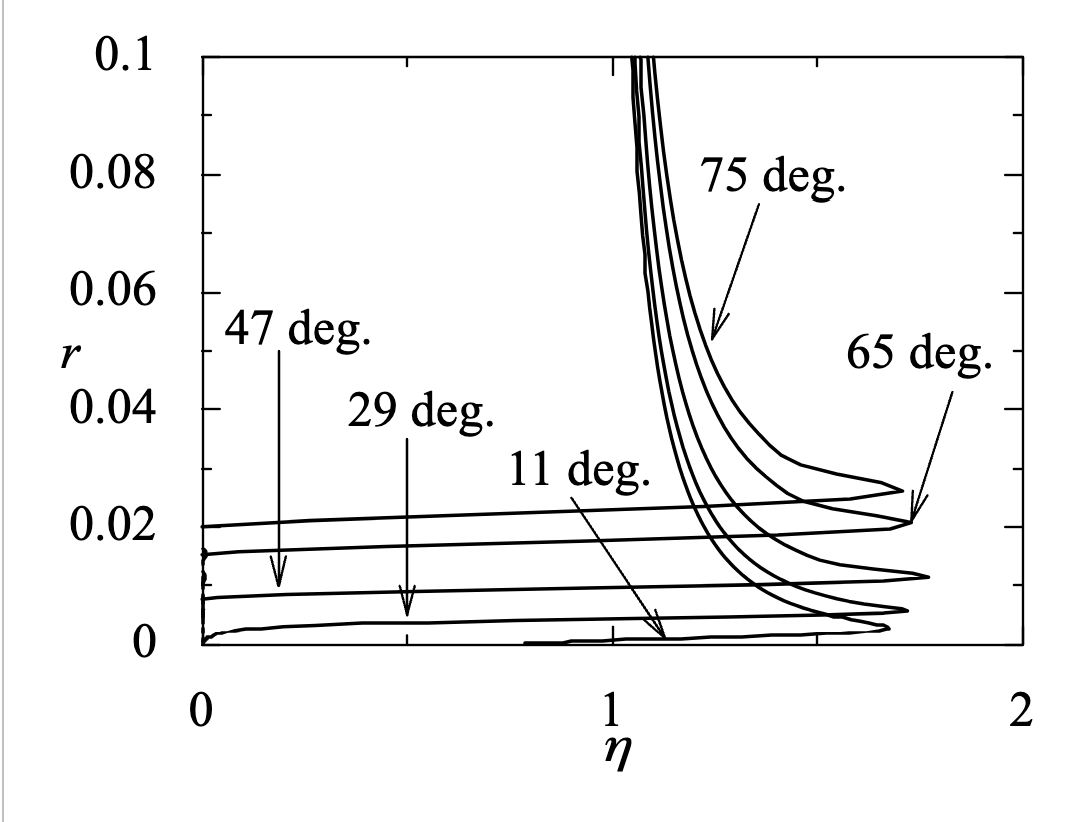}
\begin{center}
(a) Re $= 10^5$
\end{center}\label{fig:7a}
\includegraphics[width=\linewidth]{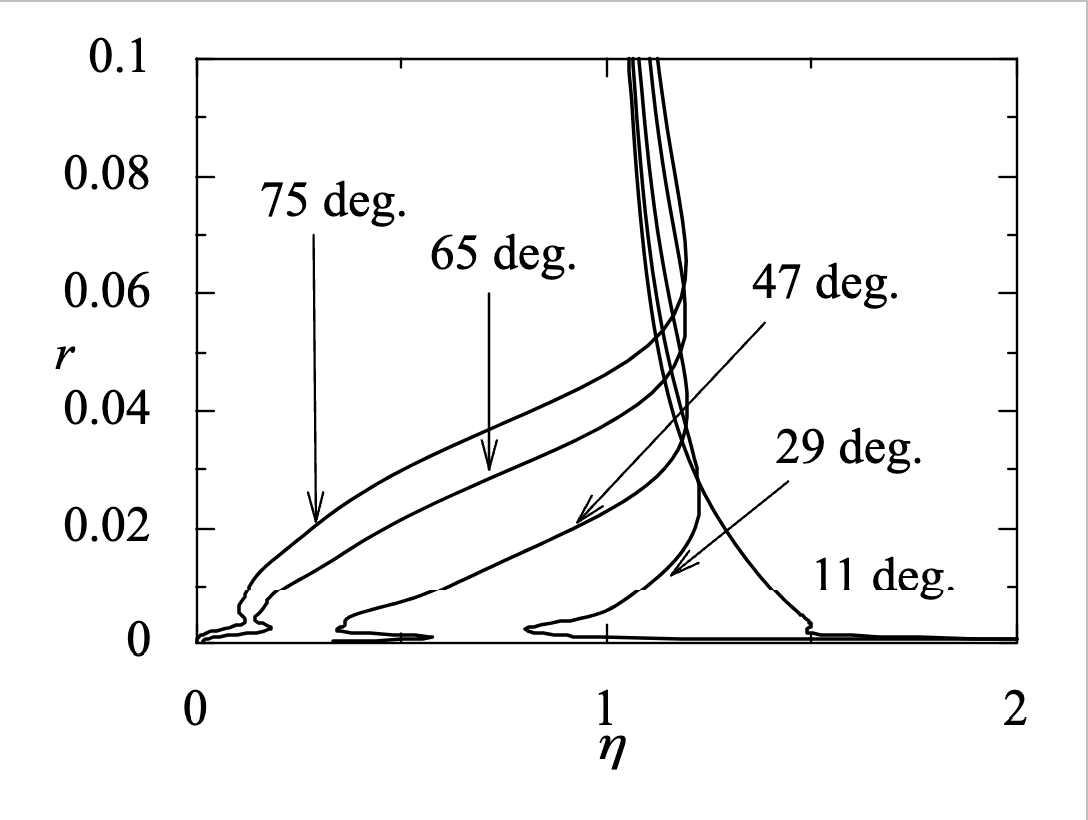}
\begin{center}
(c) Re $= 10^3$
\end{center}\label{fig:7c}
\end{minipage}
\begin{minipage}{.45\linewidth}
\includegraphics[width=\linewidth]{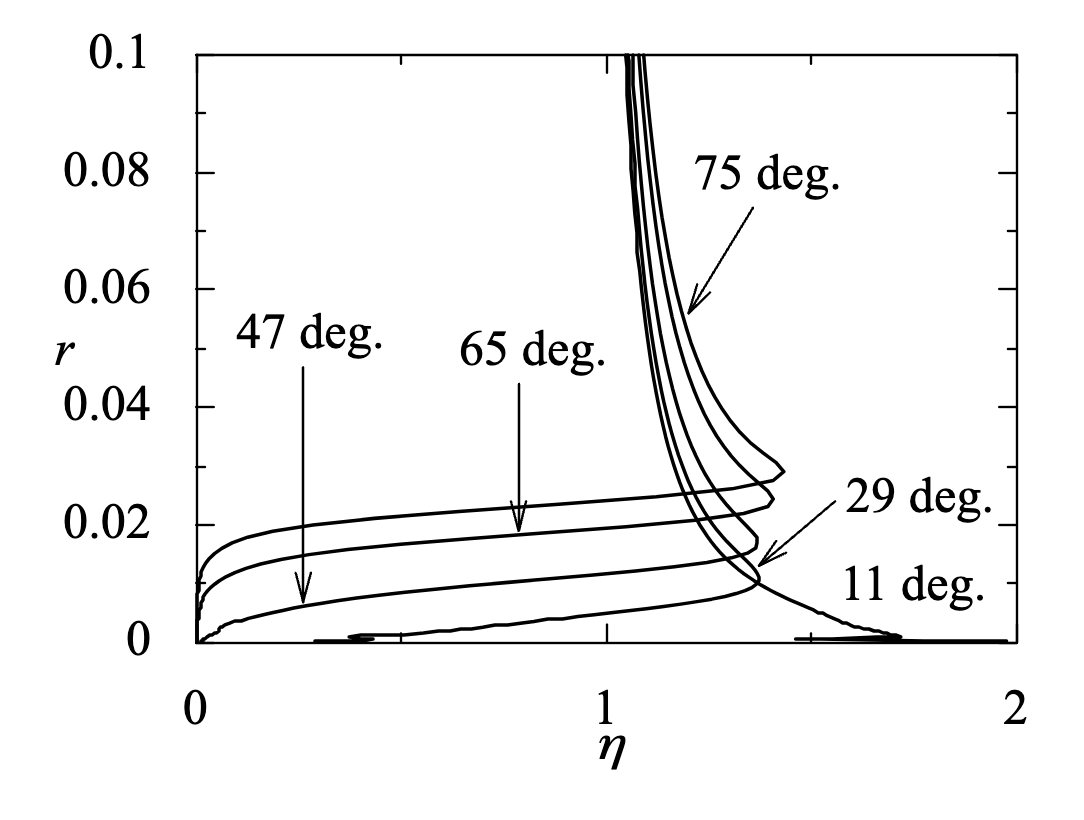}
\begin{center}
(b) Re $=10^4$
\end{center}\label{fig:7b}
\includegraphics[width=\linewidth]{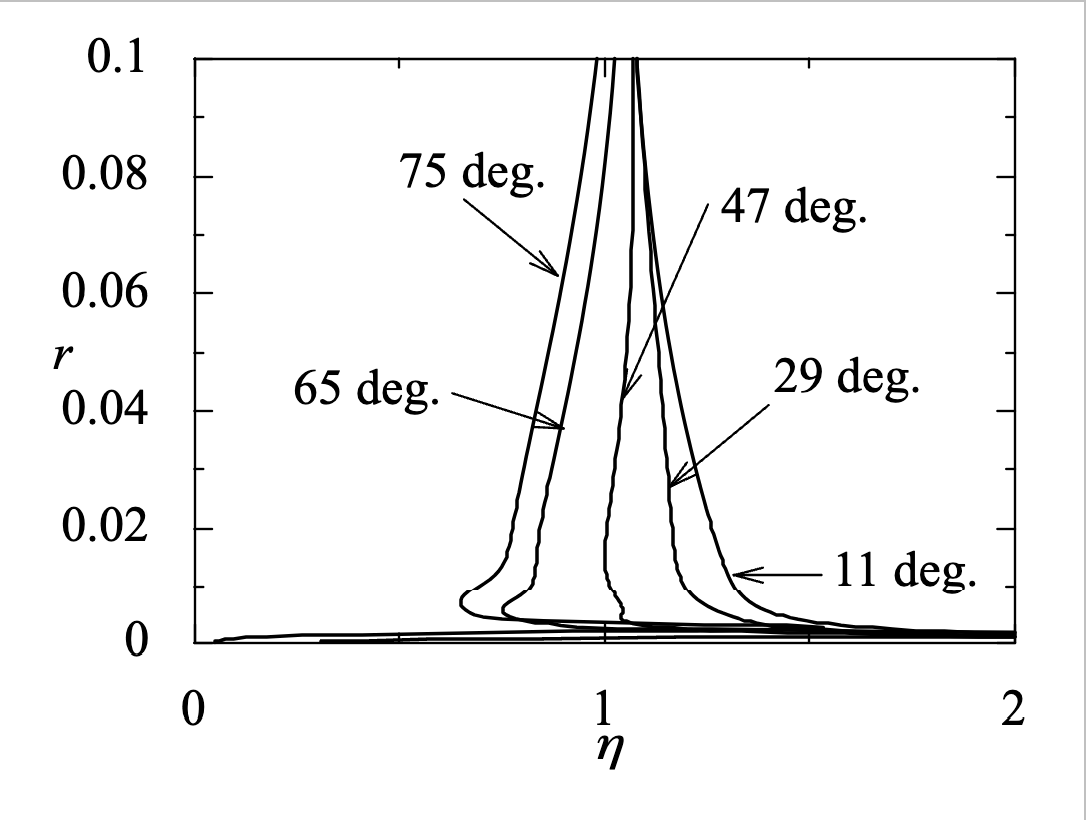}
\begin{center}
(d) Re $=10^2$
\end{center}\label{fig:7d}
\end{minipage}
\caption{Density profiles of dispersed phase along a circular cylinder ($\tau = 0.02$)}
\label{fig:7}
\end{figure}

The present computation leads to the estimation 
of the thickness of the particle-free layer as $\kappa\tau = 0.04\ (=2 \times 0.02)$. 
According to the boundary layer theory, $\mathrm{Re} \approx 600$ 
gives the corresponding thickness of the laminar boundary layer of the carrier flow. 
This implies that the particle-free layer appears when $\mathrm{Re}$ 
is approximately larger than $10^3$ in the present condition, 
which supports the criterion $\delta < \kappa\tau$ mentioned above 
and the particle-free layer appears under this condition. 
In particular, it should be noted that the particle-free layer 
is formed when $\mathrm{Re} = 10^4$. 
In this case, Michael's criterion is not satisfied 
since $\tau \mathrm{Re}^{0.5} = 2.0 \sim \mathrm{O}(1)$. 

The distribution of the thickness of the particle-free layer $\delta_p$ 
along the cylinder surface is obtained based on the results of $\mathrm{Re} \ge 10^3$. 
From the practical viewpoint, we assume that the particle-free layer forms 
under the condition $\eta \le 10^{-2}$. 
The result is shown in Figure~8, 
in which the result of the potential carrier flow is also included for $\mathrm{Re} = \infty$. 
This figure shows that the particle-free layer appears further downstream 
with decreasing the Reynolds number, 
and the thickness of the layer in the downstream region 
does not depend on the Reynolds number when $\mathrm{Re} \ge 10^4$.  

\begin{figure}[htbp]
\centering
\includegraphics[width=.45\linewidth]{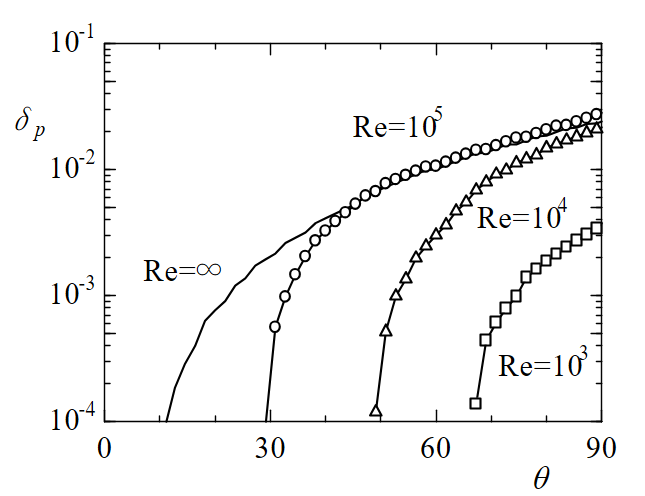}
\caption{Development of particle-free layer along a circular cylinder.}
\label{fig:8}
\end{figure}

\section{Conclusion}

We investigate some mathematical features of the density field 
of a dispersed phase in a two-fluid model of multiphase flows. 
This equation has been used commonly in icing simulations 
in order to estimate the atmospheric flow field of supercooled water droplets. 
In this study, an obstacle is placed in a uniform flow of dispersed 
and carrier phases at upstream infinity 
and the flow field of the dispersed phase around the obstacle is investigated.  

A qualitative classification of the density field of the dispersed phase 
in the stagnation region is presented for an incompressible carrier flow, 
and it is shown that the density field can be categorized 
into four types according to the Stokes number. 
Inviscid solutions in the vicinity of the stagnation point are obtained 
in exact and perturbed forms for small and large values of the Stokes number, respectively. 
A comparison between the present solutions and numerical results of the dispersed phase density exhibits good agreement, 
and the behaviour of the density is clarified over the entire range of values of the Stokes number. 
The solution shows that the density in the front stagnation region increases drastically 
when the Stokes number is close to the threshold of particle collision.  

In addition, the viscous effect of the laminar boundary layer in the carrier phase 
on the particle-free thin layer appearing in the low Stokes number regime 
is investigated based on the incompressible Navier-Stokes equations. 
The order estimation of the velocity field of the dispersed phase reveals 
the following features of this layer: 
(1) the centrifugal effect of the dispersed particles leads to the formation of the particle-free layer 
and (2) this layer appears when $\delta < \kappa \tau$. 
These conclusions are supported by numerical results. 
In particular, the second result provides an improved form of Michael's criterion for this layer.  

Finally, we note the following with regard to ice accretion simulation based on the present results. 
When the Stokes number of the icing problem under consideration is closed to 
the threshold of particle collision, the density of the dispersed phase (water droplets) 
must be estimated carefully. 
Without doing so, it is unable to estimate the incoming droplet flux on a surface correctly. 
Furthermore, when the Stokes number is smaller than the threshold, 
a singularity appears in the density field of the droplets in the stagnation region 
and a droplet free-layer forms along the downstream surface. 
In fact, it is possible for such a situation to realize locally 
if the Stokes number is close to the threshold. 
Then, the viscous effect of the airflow delays the appearance of the droplet-free layer 
and the singular region of the droplet density increases with decreasing the Reynolds number of the airflow. 
Therefore, in the computation of such small Stokes numbers, 
it is necessary to focus on the boundary conditions of the droplet density on the body surface.  

\section*{Acknowledgments}

This work was conducted while the author was affiliated with 
Major in Mechanical Systems Eng., Graduate School of Sci. and Eng., 
Ibaraki University.



\end{document}